\documentclass[prd,preprint,tightenlines,showpacs,preprintnumbers,nofootinbib,eqsecnum,superscriptaddress]{revtex4}

\usepackage{dcolumn}
\usepackage{bm}
\usepackage{epsfig}
\usepackage{amsfonts}
\usepackage{amssymb,amscd}
\usepackage{tabularx}
\usepackage{booktabs,multirow}

\def\lsim{\raise0.3ex\hbox{$<$\kern-0.75em\raise-1.1ex\hbox{$\sim$}}}

\def\gsim{\raise0.3ex\hbox{$>$\kern-0.75em\raise-1.1ex\hbox{$\sim$}}}

\newcommand{\be}{\begin{equation}}

\newcommand{\ee}{\end{equation}}

\def\beq{\begin{equation}}

\def\eeq{\end{equation}}

\def\beqa{\begin{eqnarray}}

\def\eeqa{\end{eqnarray}}

\newcommand{\ba}{\begin{eqnarray}}

\newcommand{\ea}{\end{eqnarray}}

\def\gappeq{\mathrel{\rlap {\raise.5ex\hbox{$>$}}

{\lower.5ex\hbox{$\sim$}}}}

\def\lappeq{\mathrel{\rlap{\raise.5ex\hbox{$<$}}

{\lower.5ex\hbox{$\sim$}}}}

\def\Toprel#1\over#2{\mathrel{\mathop{#2}\limits^{#1}}}

\begin{document}

\title{Probing the neutrino trident process using the Scattering and Neutrino Detector at  HL-LHC and SHiP}

\author{Reinaldo {\sc Francener}}
\email{reinaldofrancener@gmail.com}
\affiliation{Institute of Physics and Mathematics, Federal University of Pelotas (UFPel), \\
  Postal Code 354,  96010-900, Pelotas, RS, Brazil.}
\affiliation{Instituto de Física Gleb Wataghin - Universidade Estadual de Campinas (UNICAMP), \\ 13083-859, Campinas, SP, Brazil. }

\author{Victor P. {\sc Gon\c{c}alves}}
\email{barros@ufpel.edu.br}
\affiliation{Institute of Physics and Mathematics, Federal University of Pelotas (UFPel), \\
  Postal Code 354,  96010-900, Pelotas, RS, Brazil.}

\begin{abstract}
 Neutrino trident scattering is a rare process in the Standard Model characterized by two charged leptons in the final state. In this work, we investigate the possibility of probing  the neutrino trident process using the Scattering and Neutrino Detector (SND) at the Large Hadron Collider during its high - luminosity run (HL - LHC). In addition, we present, for the first time, the predictions for the neutrino trident scattering at SHiP beam - dump experiment, where a similar detector is expected to be installed. We demonstrate that these two experiments probe the process in a complementary  energy range.
Assuming the upgraded detector configuration, we estimate the cross-sections associated with all possibles leptonic final states in coherent and incoherent processes. The corresponding number of neutrino trident scatterings in the SND  at HL-LHC and SHiP are presented. Our results indicate that this process can be observed in these forthcoming experiments for some specific combinations of leptons in the final state. 
\end{abstract}



\maketitle

\vspace{1cm}

\begin{figure}[b]
	\centering
	\begin{tabular}{ccccc}
\includegraphics[width=0.48\textwidth]{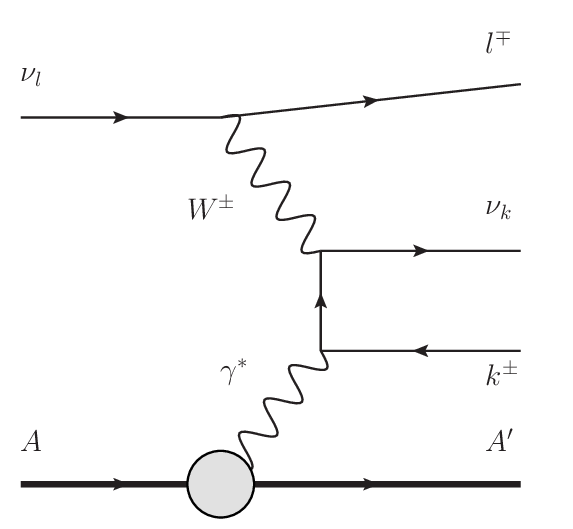} &
\includegraphics[width=0.48\textwidth]{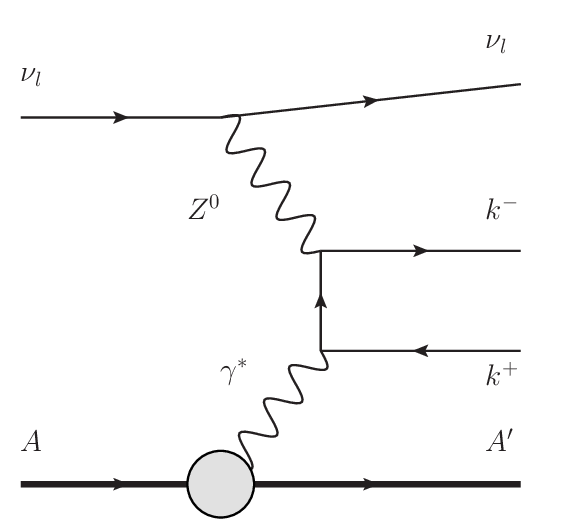} 
	\end{tabular}
\caption{ Feynman diagrams for the neutrino trident scattering off a nucleus target $A$ associated with a $W^\pm$ (left) and $Z^0$ (right) exchange. The lepton flavors $l$ and $k$ can be equal or different. }
\label{fig:diagrams}
\end{figure}

The recent discovery of collider neutrinos by the FASER \cite{FASER:2023zcr,FASER:2024hoe,FASER:2024ref} and SND@LHC \cite{SNDLHC:2023pun} collaborations has started a new physics program at the LHC, which allows testing the Standard Model (SM) and  searching for signals of New Physics in neutrino - induced interactions. Such collaborations have observed  the intense and collimated flux of neutrinos  produced in the forward direction of $pp$ collisions at the LHC, as predicted in Ref. \cite{DeRujula:1984pg}, which have motivated numerous phenomenological studies (For reviews, see e.g. Refs.\cite{Anchordoqui:2021ghd,Feng:2022inv,Anchordoqui:2026kpw}).  
  In particular, the possibility of using the  far-forward LHC detectors to probe the neutrino trident process, represented in Fig. \ref{fig:diagrams}, which is a weak process characterized by the production of a pair of charged leptons through the neutrino scattering in the Coulombian field of a heavy nucleus, was investigated in Refs. 
\cite{Francener:2024wul,Altmannshofer:2024hqd}. These studies have demonstrated that   a future observation of this rare SM process at the LHC is feasible. Such conclusion motivates the extension of these analyses for the upgraded configuration of  Scattering and Neutrino Detector (SND), recently approved to operate during the high - luminosity run of LHC (HL - LHC) \cite{SNDLHC:2026why}. This is one of our goals in this letter. The second goal is to derive, for the first time, the predictions for the number of events associated with the neutrino trident process in the SHiP experiment, which will be hosted at a new  SPS Beam Dump Facility (BDF) \cite{SHiP:2025ows}. The basic idea is that the interaction of a high - intensity proton beam with the BDF target generates an intense neutrino flux of all three flavors, which can be detected using a detector similar to the SND. As we will demonstrate below, the energy range probed by the SND@HL-LHC and SND@SHiP are complementary, with a larger number events being predicted for the SND@SHiP.



In what follows, we will present a brief review of the formalism and main assumptions present in our calculations. For a more detailed discussion we refer the interested reader to the Refs. \cite{Francener:2024wul,Altmannshofer:2024hqd} (For related studies see Refs. \cite{Magill:2016hgc,Ballett:2018uuc,Altmannshofer:2019zhy,Zhou:2019vxt,beacom2}).
We will estimate the neutrino trident process at SND@HL-LHC and SND@SHiP considering all the three neutrino flavors. The two main ingredients to evaluate the number of neutrino interactions are the neutrino-nucleus cross-section, $\sigma_{\nu A}$, and the neutrino flux reaching the detector. In our analysis, we will estimate $\sigma_{\nu A}$ using  the Monte Carlo generator developed in Ref. \cite{Altmannshofer:2019zhy} and modified in Ref. \cite{Francener:2024wul} to include the tungsten as nuclear target. Such MC generator considers the full $2\rightarrow 4$ kinematics for the trident process, without assume the validity of the equivalent photon approximation. The tungsten target is modeled with a form factor that is given by a Fourier transform of a Woods-Saxon nuclear charge distribution \cite{Woods:1954zz}, as parametrized in \cite{DeVries:1987atn}. In our calculations, we will estimate the contribution associated with  coherent and incoherent scatterings. In the coherent case, the leptonic system interacts with the full nucleus, which remains intact, which implies that there is no hadronic activity in the final state. It is important to emphasize that the coherent cross-section is proportional to the square of the nuclear charge $Z$. In contrast, in incoherent scatterings, the leptonic system interacts with the individual nucleons inside the nucleus, which implies the dissociation of the nuclear target. In this case, the cross-section contribution is proportional
to $Z$. In principle, these two contributions can be experimentally separated through the analysis of the hadronic activity in the final state.

\begin{figure}[t]
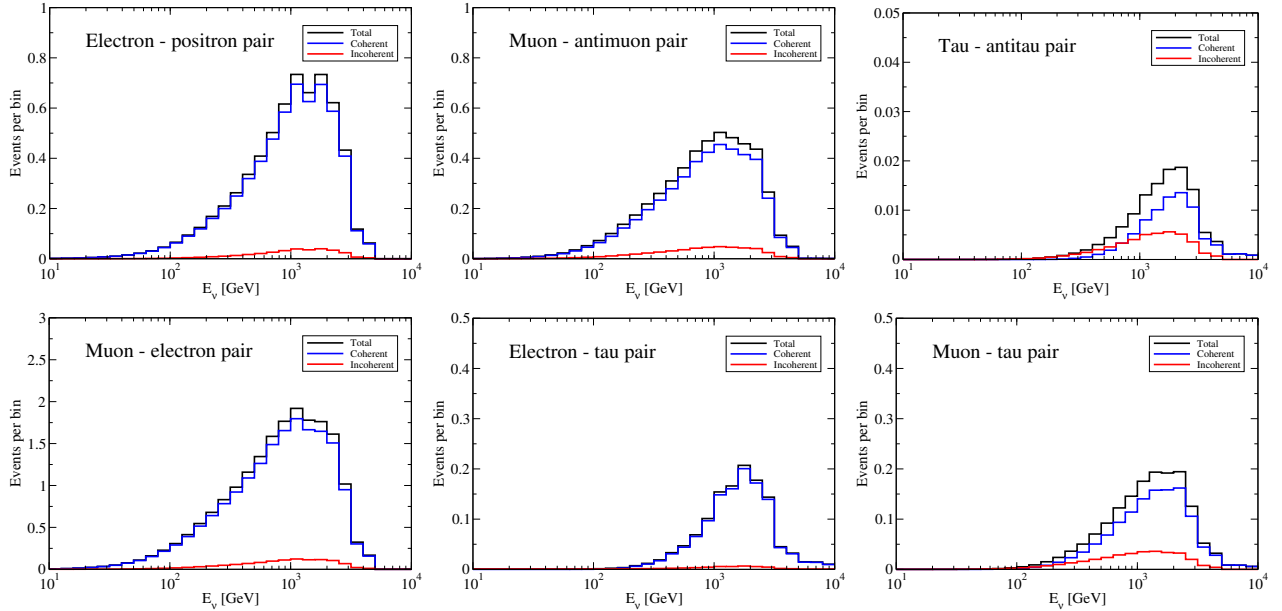

	\centering
	\begin{tabular}{ccccc}
\includegraphics[width=0.33\textwidth]{figs/Histo_events_ee_coherent_incoherent_SND.eps} &
\includegraphics[width=0.33\textwidth]{figs/Histo_events_mm_coherent_incoherent_SND.eps} & \includegraphics[width=0.33\textwidth]{figs/Histo_events_tt_coherent_incoherent_SND.eps} \\
\includegraphics[width=0.33\textwidth]{figs/Histo_events_em_coherent_incoherent_SND.eps} &
\includegraphics[width=0.33\textwidth]{figs/Histo_events_et_coherent_incoherent_SND.eps} & \includegraphics[width=0.33\textwidth]{figs/Histo_events_mt_coherent_incoherent_SND.eps} 
	\end{tabular}
\caption{ Predictions for the number of neutrino trident scattering binned in the energy of the incident neutrino at SND@HL-LHC, derived considering an integrated luminosity of 3\,ab$^{-1}$. Results for the production of  pairs of charged leptons of same (different)  flavor are presented in the upper (lower) panels. }
\label{fig:eventsSND_HL@LHC}
\end{figure}

The number of neutrino-induced events associated with the trident process is expressed in terms of the product between the time-integrated neutrino flux that reach the detector, and the probability of  interaction in this specific channel, which  is given by $\sigma_{\nu A} \rho L / m_A$, where  $\rho$ the density of the target material with nuclear mass of $m_A$, and $L$ the length of the target in the detector. In our analysis for the SND@HL-LHC,  we will assume the neutrino flux  predicted for the high luminosity LHC era presented in Ref.~\cite{Zaffaroni:2024nhf}, which was derived assuming an integrated luminosity of 3~ab$^{-1}$.
On the other hand, for the  SND@SHiP, we will consider the neutrino flux derived in Ref.~\cite{SHiP2023}, which assumes that CERN SPS accelerator can deliver $6 \times 10^{20}$ protons on target (PoT) of 400~GeV during 15 years. Regarding the detector configuration for the SND@HL-LHC, we will consider the extended configuration discussed in Ref.~\cite{SNDLHC:2026why}, which assume the tungsten as a target composed by 58 plates of 7~mm each. In contrast, for the SND@SHiP, we will assume  the most recent version presented in Ref.~\cite{SHiP2025}, which proposed a composed detector with 40~cm$\times$40~cm of transverse area. The first part of the detector is composed of 120 tungsten ($\rho = 19.3$~g/cm$^{3}$) plates with thickness of 3.5~mm, totalizing $\approx$ 1.3 metric tons of tungsten. The second part has iron ($\rho = 7.87$~g/cm$^{3}$) as the target, with 42 plates of 5~cm of thickness and a total of $\approx$ 2.6 metric tons.

Initially, in Fig.~\ref{fig:eventsSND_HL@LHC}, we present our results for the number of events associated with the neutrino trident process at the SND@HL-LHC, binned in the energy of the incident neutrino.  The predictions for coherent and incoherent interactions, as well for the sum of both, are shown for different combinations of charged leptons in the final state. Our results indicate that the peak of the distribution occurs at TeV energies, which is expected, given the characteristic of the incoming neutrino flux produced in $pp$ collisions at $\sqrt{s} = 14$ TeV~\cite{Zaffaroni:2024nhf}. Moreover, our results indicate that the coherent process dominates, with the relative contribution from coherent interactions being dependent on the final state considered. Such conclusion is confirmed in Table~\ref{table:Nevents_HL-LHC}, where we present the total number of events at SND@HL-LHC. Moreover, these results also demonstrate that the larger event rate is associated with the $e^{\pm} + \mu^\mp$ final state. A similar dominance was obtained in Ref.~\cite{Francener:2024wul}, where the total number trident events in the FASER$\nu$2 detector at HL-LHC was estimated. In comparison with the results presented in Ref.~\cite{Francener:2024wul}, we predict that the number of events at SND@HL-LHC is approximately one order of magnitude smaller than at FASER$\nu$2. However, our results indicate that the probing of the neutrino trident process at SND@HL-LHC is still feasible.

\begin{table}[t]
\centering
\renewcommand{\arraystretch}{1.5}
\begin{tabularx}{\textwidth}{XXXl}
\toprule
\hline
\multicolumn{4}{c}{\bf Events for neutrino trident scattering at SND@HL-LHC}\\
\hline
\midrule
{\bf Final state}      & {\bf Incoherent} & {\bf Coherent} & {\bf Total} \\
\hline	
\toprule   
$e^{+} + e^-$          &    $0.33$       &   $5.97$     & $6.30$ \\
\hline
\midrule
$\mu^{+} + \mu^-$      &     $0.49$      &    $4.48$     & $4.97$ \\
\hline
\midrule
$\tau^{+} + \tau^-$    &     $0.04$       &    $0.08$    & $0.12$ \\ 
\hline   
\midrule
$e^{\pm} + \mu^\mp$    &     $1.17$      &    $17.73$    & $18.90$ \\ 
\hline
\midrule
$e^{\pm} + \tau^\mp$   &    $0.06$        &    $1.21$      & $1.27$ \\  
\hline
\midrule
$\mu^{\pm} + \tau^\mp$ &    $0.33$        &    $1.23$      & $1.56$ \\ 
\hline
\bottomrule
\end{tabularx}
\vspace{0.3cm}
\caption{ Predictions for the number of neutrino trident scattering at the SND@HL-LHC assuming an integrated luminosity of 3 ab$^{-1}$. We are summing over neutrino, antineutrino and three flavors  of the incoming neutrinos. Results for coherent and incoherent interactions and distinct leptonic final states.
}
\label{table:Nevents_HL-LHC}
\end{table}

\begin{figure}[t]
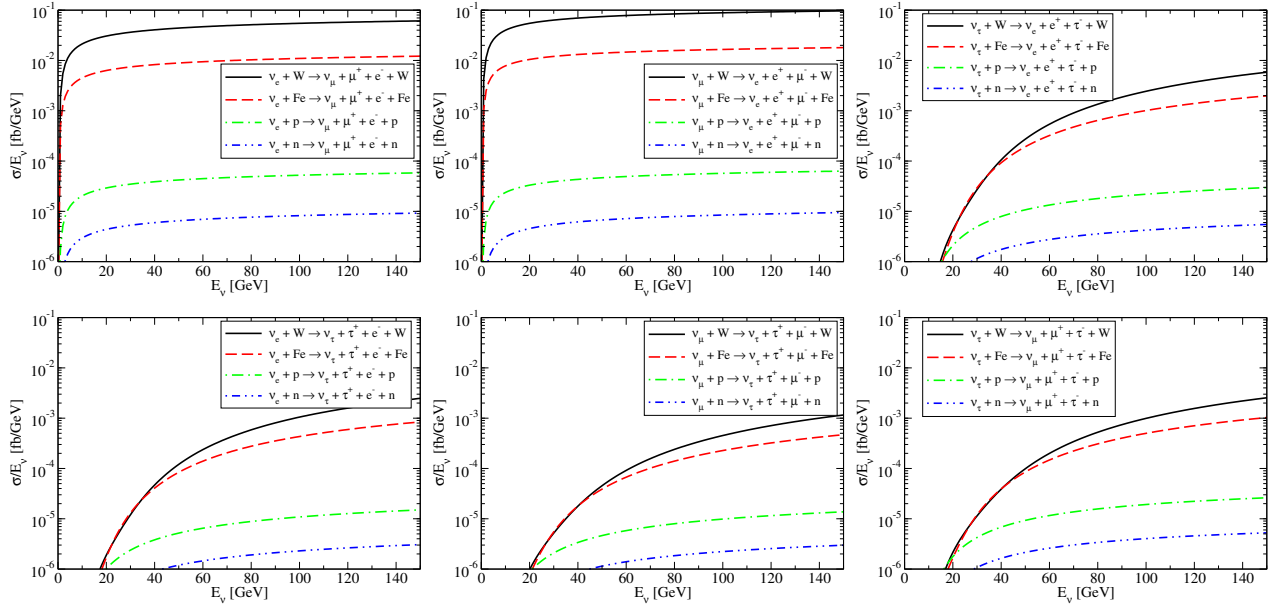

	\centering
	\begin{tabular}{ccccc}
\includegraphics[width=0.33\textwidth]{figs/cs_nue_em.eps} &
\includegraphics[width=0.33\textwidth]{figs/cs_num_em.eps} & \includegraphics[width=0.33\textwidth]{figs/cs_nut_et.eps} \\
\includegraphics[width=0.33\textwidth]{figs/cs_nue_et.eps} &
\includegraphics[width=0.33\textwidth]{figs/cs_num_mt.eps} & \includegraphics[width=0.33\textwidth]{figs/cs_nut_mt.eps} 
	\end{tabular}
\caption{ Energy dependence of the cross-section, normalized by $E_\nu$,  for the neutrino trident process. Results for the production of different charged leptonic flavors, derived assuming an incident  electronic (left panel), muonic (center panel) and tauonic (right panel) neutrino. The predictions  are shown for  coherent scattering with tungsten (continuous black line) and iron (dashed red line), and incoherent scattering with protons (dot-dashed green line) and neutrons (dot-dashed blue line). }
\label{fig:CS_energy}
\end{figure}

Differently from the neutrino flux generated in $pp$ collisions at the LHC, which is characterized by neutrinos with energies of order of TeV, the fixed - target proton - nucleus collisions at SHiP produce a flux of neutrinos with $E_{\nu}$ of order of few dozen of GeV. As a consequence, the  neutrino trident events at SND@SHiP are expected to be determined by $\sigma_{\nu A}$ at smaller energies than in the SND@HL-LHC. For completeness of our analysis, in Fig.~\ref{fig:CS_energy} we present the energy dependence of the coherent and incoherent cross-sections in the energy range of interest, considering the production of a charged pair of different leptonic flavors. As the SND@SHiP is composed by tungsten and iron nuclear targets, we present the predictions for the coherent cross-section for these two nuclei. We also show the results for incoherent interaction with a proton and neutron target. The cross-section for the incoherent scattering with a proton is approximately one order of magnitude larger than for a neutron target, which is smaller since it is electrically neutral. Moreover, we have that the cross-section for the coherent interaction with a tungsten is larger than with an iron target, which is expected since the coherent cross-section is proportional to $Z^2$. Another important aspect is that the cross-section decreases for larger masses of the charged leptonic system, and that the presence of a tau in the final state implies that the cross-section  becomes appreciable only for $E_{\nu} \gtrsim 20$ GeV.

\begin{figure}[t]
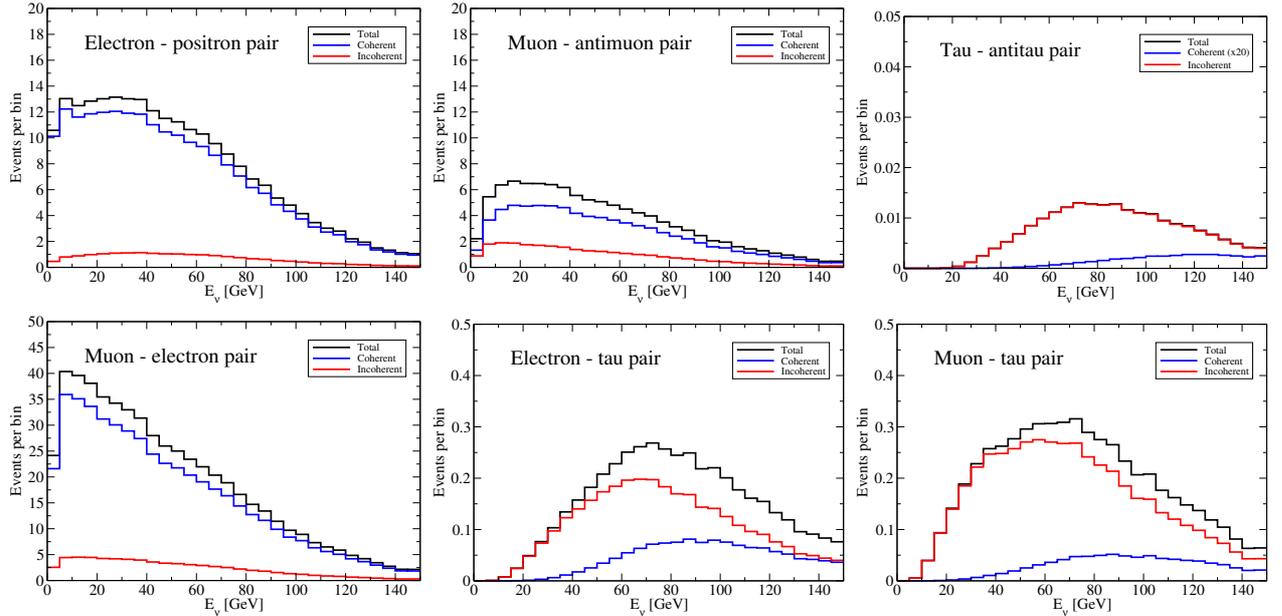

	\centering
	\begin{tabular}{ccccc}
\includegraphics[width=0.33\textwidth]{figs/Histo_events_ee_coherent_incoherent.eps} &
\includegraphics[width=0.33\textwidth]{figs/Histo_events_mm_coherent_incoherent.eps} & \includegraphics[width=0.33\textwidth]{figs/Histo_events_tt_coherent_incoherent.eps} \\
\includegraphics[width=0.33\textwidth]{figs/Histo_events_em_coherent_incoherent.eps} &
\includegraphics[width=0.33\textwidth]{figs/Histo_events_et_coherent_incoherent.eps} & \includegraphics[width=0.33\textwidth]{figs/Histo_events_mt_coherent_incoherent.eps} 
	\end{tabular}
\caption{ Predictions for the number of neutrino trident scattering binned in the energy of the incident neutrino at SND@SHiP during 15 years. The results for pairs of charged leptons of same (different) flavors are presented in the upper (bottom) panels. }
\label{fig:eventsSHiP}
\end{figure}

In Fig.~\ref{fig:eventsSHiP} we present our predictions for the number of neutrino trident scattering binned in the energy of the incident neutrino at SND@SHiP during 15 years, with  the results for the production of  pairs of charged leptons of same (different)  flavor are presented in the upper (lower) panels. We have that the peak of the distributions occurs for $E_{\nu} \approx 20$ GeV when a pair of light leptons is produced. On the other hand, if a tau is present in the final state, the peak occurs for larger neutrino energy.  Another important aspect is that the incoherent contribution becomes larger with the increasing of the mass of the leptonic system, being dominant when a tau lepton is produced, in contrast to observed at SND@HL-LHC. Such conclusions are confirmed by the results presented in Table~\ref{table:Nevents_SND@SHiP}, where the number of events associated with the neutrino scattering process at SND@SHiP are presented. The calculations have been performed considering the two nuclear targets, with the number of events for coherent interactions for each nuclear target being presented separately. For incoherent interactions, we only are presenting the total contribution, considering the scattering with protons and neutrons in the distinct targets. We have that the $e^{\pm} + \mu^\mp$  final state has the larger event rate, as in the SND@HL-LHC. Our results indicate that the event rate {\it per year} at SND@SHiP associated with the $e^+ + e^-$, $\mu^+ + \mu^-$ and $e^{\pm} + \mu^\mp$  final states are a factor $\approx$ 2 larger than those predicted for the  SND@HL-LHC. In contrast, the analysis of final states where a tau lepton is produced will be a hard task at SND@SHiP.   

\begin{table}[t]
\centering
\renewcommand{\arraystretch}{1.5}
\begin{tabularx}{\textwidth}{XXXXc}
\toprule
\hline
\multicolumn{5}{c}{\bf Events for neutrino trident scattering at SND@SHiP}\\
\hline
\midrule
{\bf Final state}      &  {\bf Incoherent} & {\bf Coherent ($W$)} & {\bf Coherent ($Fe$)} & {\bf Total} \\
\hline	
\toprule    
$e^{+} + e^-$          &      $19.69$   & $153.96$ &   $55.28$     & $228.93$ \\
\hline
\midrule
$\mu^{+} + \mu^-$      &      $26.75$   & $52.14$  &    $25.28$     & $104.17$ \\
\hline
\midrule
$\tau^{+} + \tau^-$    &      $0.20$   & $6.46 \times 10^{-4}$ &    $1.27 \times 10^{-3}$    & $0.20$ \\ 
\hline   
\midrule
$e^{\pm} + \mu^\mp$    &      $69.20$   & $347.64$ &    $136.97$    & $553.81$ \\ 
\hline
\midrule
$e^{\pm} + \tau^\mp$   &     $3.17$    & $0.66$    &    $0.60$      & $4.43$ \\  
\hline
\midrule
$\mu^{\pm} + \tau^\mp$ &     $4.73$    & $0.39$    &    $0.43$      & $5.55$ \\ 
\hline
\bottomrule
\end{tabularx}
\vspace{0.3cm}
\caption{ Predictions for the number of neutrino trident scattering at the SND@SHiP considering 15 years of data collection and the two nuclear targets. We are summing over neutrino, antineutrino and three flavor of the incoming neutrinos. Results for coherent and incoherent interactions and distinct leptonic final states. 
}
\label{table:Nevents_SND@SHiP}
\end{table}

In summary, in this letter we have investigated the possibility of probing  the neutrino trident process using the forthcoming Scattering and Neutrino Detectors to be installed  at LHC during the high luminosity run and in the SHiP experiment at the  new SPS Beam Dump Facility. Considering the proposed detector configurations, we have estimated the event rates associated with the neutrino trident process in coherent and incoherent interactions. Our results indicate that the analysis of this process is, in principle, feasible in both experiments, especially if the $e^{\pm} + \mu^\mp$  final state is considered, with the event rate / year for the production of a pair of light charged leptons at SND@SHiP being a factor $\approx 2$ larger. We expect that our results motivate a future experimental analysis of the neutrino trident scattering in these forthcoming experiments.

\begin{acknowledgments}
The work of R.F. was supported by Ministry of Science, Technology and Innovation - MCTI, National Council for Scientific and Technological Development – CNPq (Process No. 161770/2022-3) and FAPERGS. VPG was partially financed by the Brazilian funding agencies CNPq, FAPERGS and INCT-FNA  (Process No. 408419/2024-5). 
\end{acknowledgments}

\end{document}